# Threat-Informed Cyber Resilience Index:
# A Probabilistic Quantitative Approach to Measure Defence Effectiveness Against Cyber Attacks

Lampis Alevizos
*Volvo Group*
*Amsterdam, The Netherlands*
lampis@redisni.org

Vinh- Thong Ta
*Edge Hill University*
*Ormskirk, UK*
tav@edgehill.ac.uk

*Abstract — In the dynamic cyber threat landscape, effective decision-making under uncertainty is crucial for maintaining robust information security. This paper introduces the Threat-Informed Cyber Resilience Index (CRI), a probabilistic approach to quantifying an organisation's defence effectiveness against cyberattacks (attack campaigns). Building upon the Threat-Intelligence Based Security Assessment (TIBSA) methodology, we present a mathematical model that translates complex threat intelligence into an actionable, unified metric similar to a stock market index, that executives can understand and interact with while teams can act upon. Our method leverages Partially Observable Markov Decision Processes (POMDPs) to simulate attacker behaviour considering real-world uncertainties and the latest threat actor tactics, techniques, and procedures (TTPs). This allows for dynamic, context-aware evaluation of an organisation's security posture. As a result, decision-makers are equipped with a single metric of cyber resilience that bridges the gap between quantitative and qualitative assessments, enabling data-driven resource allocation and strategic planning. This can ultimately lead to more informed decision-making, mitigate under or overspending, and assist in resource allocation.*

## I. Introduction

The rapid evolution of the information security domain necessitates innovative approaches to manage uncertainty effectively. Traditional risk assessment methodologies often fall short of addressing the dynamic nature of cyber threats. The inclusion of cyber threat intelligence (CTI) provides a richer context for understanding potential threats, yet there remains a need for a comprehensive, quantifiable measure of an organization's security posture. This paper enhances the Threat-Intelligence Based Security Assessment (TIBSA) methodology [1] with a unified metric that we coin as the Cyber Resilience Index (CRI). The CRI serves as a unified metric to evaluate and enhance the cyber resilience of organizations, providing a clear, actionable framework for decision-makers.

The CRI functions similarly to financial indices such as the S&P 500, offering a daily metric that executives can use to steer their defences, regulate budgets, and adjust processes and investments based on risk tolerance and risk appetite. Security teams can also act on the CRI to enhance their defences, making it a practical tool for continuous improvement in cybersecurity.

The concept of uncertainty management in information security has been increasingly emphasized, particularly with the updates in risk definitions within standards such as ISO 31000:2018 [2] and ISO 27005:2022 [3]. These standards recognize the inherent unpredictability of cyber threats and the necessity for flexible and adaptive risk management strategies. Traditional risk assessment frameworks, including ISO 27005 and NIST CSF 2.0 [4], offer comprehensive approaches but often lack mechanisms to adequately incorporate and address uncertainty.

Cyber threat intelligence (CTI) has emerged as a critical component in reducing this uncertainty by providing actionable insights into potential threats. Despite the advances in CTI, a gap remains in translating these insights into a unified, quantifiable metric that can guide decision-making processes effectively. This paper aims to fill this gap by introducing the CRI as part of the TIBSA methodology, offering a robust framework for evaluating and enhancing cyber resilience.

## II. Literature Review

Traditional risk assessment methodologies, such as ISO 27005 [4], NIST SP 800-30 [5], IRAM2, and others provide structured approaches to identify, assess, and mitigate risks. These frameworks typically involve qualitative, quantitative, and semi-quantitative methods to evaluate the effectiveness of security controls. While comprehensive, these methodologies are not agile enough to address the rapidly evolving cyber threats and the uncertainty inherent in modern cyber environments. For instance, qualitative methods may rely heavily on expert judgment, which can introduce bias and subjectivity [5], [6]. Quantitative methods on the other hand, while more objective, may fail to capture the full spectrum of risks due to limitations in data availability and modelling techniques [7], [8].

CTI provides critical insights into threat actors, their tactics, techniques, and procedures (TTPs), and the potential impact on organizational assets [9]. Studies have shown that incorporating CTI into risk management processes can significantly enhance the ability to predict and mitigate cyber threats [1]. However, the integration of CTI into a cohesive, actionable metric remains a challenge. Existing CTI frameworks, such as MITRE ATT&CK [10], offer valuable information on adversary behaviour but often require significant expertise to interpret and apply effectively.

Various metrics and models have been proposed to quantify cyber resilience, but there remains a need for a unified, comprehensive measure. Several scholars have made significant contributions to this field, for instance, the cyber resilience review (CRR) developed by the U.S. Department of Homeland Security, is a non-technical assessment to evaluate an organization's operational resilience and cybersecurity practices. While comprehensive, it is largely qualitative and can be subjective, relying heavily on self-assessment [11]. The CRR's lack of quantitative rigour means it can be challenging to track improvements over time or compare results across different organizations objectively. In contrast, the CRI provides a continuous, quantifiable metric that can be easily tracked and benchmarked.

The Cybersecurity Capability Maturity Model (C2M2) assesses the maturity of an organization's cybersecurity capabilities. It is useful for identifying gaps and areas for improvement but does not provide a unified metric that can be easily tracked over time [12]. C2M2 focuses more on process maturity rather than direct measurement of resilience against threats. The CRI, on the other hand, offers a real-time, actionable score that reflects the current effectiveness of security measures against the latest threats, providing a dynamic and responsive tool for decision-makers.

Bodeau et al. proposed a set of metrics to assess cyber resilience, focusing on the ability to anticipate, withstand, recover, and adapt to cyber incidents [13]. These metrics are valuable but can be complex to implement and require detailed data collection and analysis, which may not be feasible for all organizations. Additionally, the individual metrics do not coalesce into a single, unified score. The CRI simplifies this complexity by aggregating various resilience factors into a single, easy-to-understand index, making it more accessible and practical for executives and security teams alike.

Petit et al. proposed the resilience measurement index (RMI) which aims to provide a quantifiable measure of resilience based on factors such as robustness, resourcefulness, and adaptability [16]. While promising, the RMI lacks widespread adoption, it does not consider cyber threat intelligence as a primary source, and it has become outdated since publishing. Its comprehensive nature can also make it difficult to implement and interpret without significant expertise. The CRI builds on the conceptual strengths of the RMI but offers a more streamlined and widely applicable approach, backed by real-world implementation through the TIBSA methodology.

World Economic Forum (WEF) in collaboration with Accenture proposed a cyber resilience framework to evaluate the resilience of national and organizational cyber defences [14]. It focuses on strategic, operational, and governance aspects of cyber resilience, and while it offers a high-level overview, it is less actionable at the operational level due to its broad scope. The CRI proposed in this paper focuses specifically on operational effectiveness and actionable insights for organizations, providing a detailed and dynamic metric that can be used for continuous improvement and immediate decision-making.

The concept of cyber resilience extends beyond traditional risk management to incorporate the ability of an organization to anticipate, withstand, and recover from cyber incidents. As the literature reveals, various metrics and models have been proposed to quantify cyber resilience, but there remains a need for a unified, comprehensive measure. The Cyber Resilience Index (CRI) proposed in this paper aims to fill this gap by providing a single, quantifiable score that reflects the effectiveness of security controls validated through the TIBSA methodology.

The CRI is designed to function much like a stock market index, offering a clear and actionable metric that executives can monitor regularly. As new threat actors are identified through CTI, organizations can assess, validate, and measure their defences against these threats, updating the CRI accordingly. This enables decision-makers to allocate resources, adjust policies, and improve processes based on real-time assessments of their cyber resilience, always aligned with their risk tolerance and appetite.

Finally, we discuss some relevant related works that applied the Partially Observable Markov Decision Process (POMDP) to model cybersecurity problems. The authors in [15] formulate a Partially Observable Markov Decision Process (POMDP) to optimise Moving Target Defence (MTD) strategies in cybersecurity. The authors propose a model that balances unpredictability and system manageability by abstracting defence priorities into a POMDP. An autonomous agent uses this model to select optimal defensive actions based on the assessed phase of a cyber-attack. The model achieves high attack suppression (greater than 99%) and system availability (greater than 94%) even with partial detection capabilities. Yue et al. [16] introduce the ND3RQN algorithm that integrates Long Short-Term Memory (LSTM) to manage historical data and improve decision-making. The goal of the paper is to enhance the efficiency of automated penetration testing by modelling the process as a POMDP and improving the Deep Recurrent Q-Network (DRQN) algorithm. The ND3RQN algorithm offers superior performance in discovering attack paths and potential vulnerabilities without prior information about the network. In [17], the authors addressed the problem of a scalable defence strategy for large-scale cyber networks using POMDP to mitigate adversary progression in real-time. The model's scalability might be constrained by the complexity of maintaining accurate belief states and processing large volumes of security alerts in real-time. Wang et al. [18] model and analyse honeypot systems using both Markov Decision Processes (MDP) and Partially Observable Markov Decision Processes (POMDP). It explores different analytical methods including value iteration, policy iteration, linear programming, and Q-learning for MDP, and applies these to honeypot systems. It also examines the effects of observation probabilities and rewards in a POMDP framework. Finally, Sarraute et al. [19] propose modelling penetration testing as a POMDP problem, integrating information gathering with exploit planning. Their approach addresses the uncertainties in network configurations and probabilistic outcomes of scanning and exploitation actions. This model aims to optimize attack strategies under uncertainty by intelligently mixing scanning actions with exploits. The study highlights the importance of considering residual uncertainties and the cost-effectiveness of scanning actions. However, the scalability of the model to large networks remains a challenge, and practical implementation may require significant computational resources.

## III. Proposed CRI Calculation Framework

Given an attack campaign such as the Lockbit ransomware campaign [20], WannaCry ransomware [21], Marriott Data Breach [22], Yahoo Data Breach [23], and Operation Aurora [24], our framework aims to provide a quantitative cyber resilience metric (CRI) of a given computer system or network against these attacks.

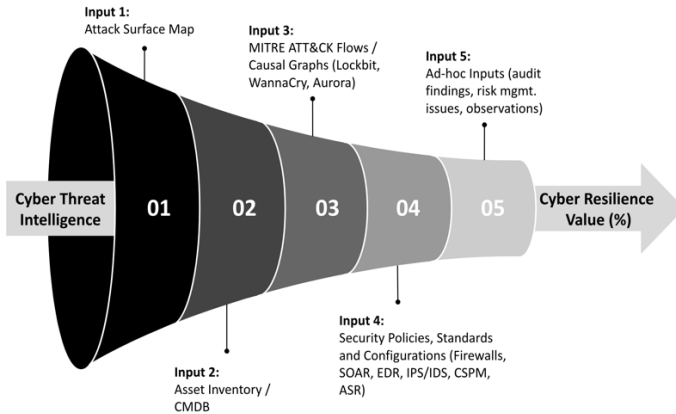

*Figure 1. Cyber Resilience Index (CRI) framework.*

Generally speaking, to calculate accurately the resilience or defence effectiveness of a computer system/network against an attack campaign, five inputs are typically required (as can be seen in Figure: (1) the attack surface of an organisation network/system (2) the asset inventory or configuration management database extract (3) threat intelligence regarding cyber threats, and the modus operandi of the corresponding threat actors, and (4) a set of security policies, standards, configurations from security telemetry and returns a cyber resilience metric in the form of statistical values. The fifth input (5) is reserved for ad-hoc measurements, such as resolved audit findings, mitigated issues, or red team exercises that would positively or negatively impact the final resilience index. Threat intelligence data is used by our framework to estimate factors such as the attack frequency, success probability of a typical technique as well as the effectiveness of countermeasures in place.

Our framework addresses these, and for mathematical calculation purposes, we specify these inputs as follows: The attack surface is captured/modelled as part of an organisation system/network. The asset inventory is defined for the devices in the network including the attack surfaces, while the security policies are specified on the entire network. The threat intelligence regarding cyber threats in our framework is captured with known cyber-attack campaigns and is modelled with a set of MITRE ATT&CK attack flows (we will discuss this in detail in Section IV.). Due to complexity, we leave the consideration of the fifth input for future work and focus on the first four.

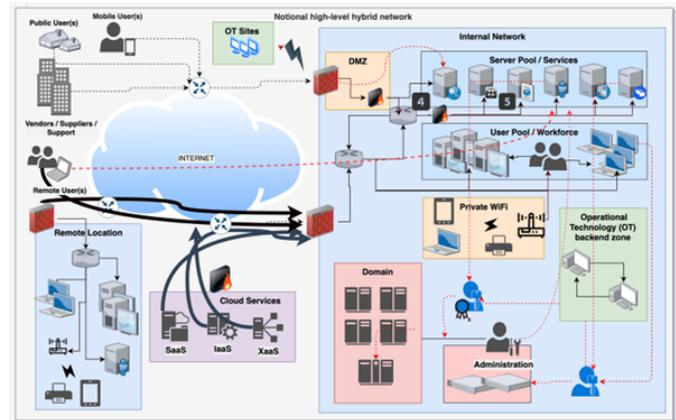

*Figure 2. A typical corporate network, which is still universally applied nowadays.*

A typical network is depicted in Figure 2. This illustrates a high-level overview of a notional company's hybrid network, integrating Information Technology (IT) and Operational Technology (OT) systems. The notional network is designed to support a variety of user types and connection methods. The internal network is the core of the company's IT infrastructure, where we see a pool of servers providing diverse services. Alongside the servers, there is a user pool representing the workforce-connected endpoints. A Demilitarized Zone (DMZ) is a secure area that sits between the internal network and the external internet, designed to provide an additional security layer by hosting the services that are accessible from the outside. The (optional) Operational Technology (OT) backend zone is part of the network dedicated to managing and monitoring industrial control systems and other equipment crucial for operational tasks. Administration is a designated section for network administrators, implying centralized control over network operations. Private WIFI denotes a wireless network within the company, typically used for internal communications and mobile device connectivity. Domain controllers' zone is the area marked as the domain, indicating a collection of networked computers providing typical intra-domain services and security policies. Cloud Services represent Software as a Service (SaaS), Infrastructure as a Service (IaaS), and other cloud-based services (XaaS), showing the company's use of external resources. Remote location signifies a branch or separate facility of the company that is connected to the main network, through a typical virtual private network (VPN) channel. There is also a remote user and mobile users' group, accessing the network from various locations, outside of the company's premises. This group may also represent customers or employees on the go. Vendors, suppliers, and supports indicate external parties that interact with the company's network, typically for maintenance or supply chain activities. The entire network lacks internal segregation, which suggests that the company has a 'flat network' with no additional security barriers between the different internal sections. The design facilitates connectivity and resource sharing but may require robust security protocols to protect against internal and external threats.

## IV. Specification and Modelling of Inputs

In our framework, an attack campaign is provided by CTI in the form of a structured report detailing the threat actor's modus operandi, modelled and specified by a set of TTPs (tactics, techniques, and procedures) using attack flows [25]. A computer system/network is modelled by a causal graph $G= ((V, I), E)$, where $(V, I)$ is a pair of a node and inventory set attached to this node, respectively. A node represents an asset such as a server, an endpoint, or a networking device (e.g., firewall, IDS, IPS). The inventory set consists of software, apps, and services running inside a node. An edge connecting two nodes represents their network connection. The threat intelligence-related parameters and arguments are numerical and statistical values. Finally, the security policies are specified in policy languages (access policy language, firewall and IDS rules specification language, and a language for specifying network segmentation policies).

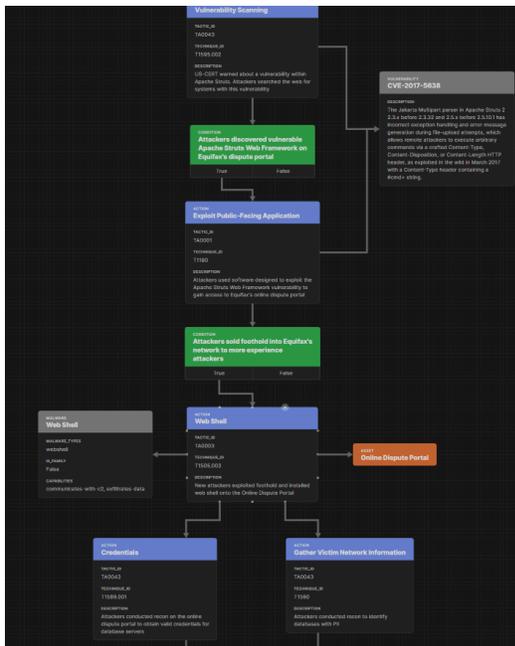

*Figure 3. An attack flow segment of the infamous Equifax Breach* [26].

In the following, we will discuss the specification of each input in more detail. The TTPs (Tactics, Techniques, and Procedures) attack flow by MITRE outlines a structured approach for understanding and documenting cyber-attack behaviour. It focuses on categorizing and describing how adversaries conduct attacks, including their overarching strategies (tactics), specific methods used (techniques), and detailed implementation details (procedures). Industry uses MITRE's TTP (Tactics, Techniques, and Procedures) framework to systematically describe and understand cyber-attack campaigns by mapping observed attack behaviours to standardized categories. This helps in identifying threat patterns, enhancing detection and response capabilities, and facilitating communication of threat intelligence across organizations.

As shown in Figure 3, an attack flow is built up of nodes and edges, where the nodes specify the:

- Tactics: High-level objectives or goals that the attacker aims to achieve (e.g., Initial Access, Execution).
- Techniques: Specific methods or actions used to achieve the tactics (e.g., Phishing, Command and Scripting Interpreter).
- Procedures: Detailed, concrete implementations of techniques that are used in specific attack scenarios.

While the edges specify the following:

- Flow Relationships: Arrows or lines connecting nodes to illustrate the sequence and progression of actions from one tactic or technique to another.
- Dependencies and Causal Links: Indicate how one action enables or influences subsequent actions, showing the logical flow of the attack.

Regarding the computer system/network, we provide an example related to the notional network in Figure 2. The set of nodes in this case can be defined by V, where {PeriFw, VPNGw, DMZFw, IntRouter, DMZRouter, IDS, StaffEndPoint, AdminEndPoint, StaffRemoteEndPoint, MailServer, WebServer, FileServer} ⊆ V, with the perimeter firewall, VPN gateway, DMZ firewall, internal router/switch, DMZ router, intrusion detection system, staff endpoints and admin endpoints, and remote endpoints of staff, then the three servers in the server pool. Each node above has a corresponding inventory set that specifies the concrete hardware, firmware, and software parameters including operating systems, apps, and services running in them.

In our example, the set of edges is defined by E, where {(PeriFw, VPNGw), (VPNGw, DMZFw), (IntRouter, DMZRouter), (DMZRouter, IDS), (IDS, StaffEndPoint), (IDS,StaffEndPoint), (IDS, AdminEndPoint), (IDS, MailServer), (IDS, WebServer), (IDS, FileServer)} ⊆ E. Each edge is defined by a pair of nodes it connects. Edges represent physical connections between hardware assets. Based on the edges, we can identify the physical paths that connect two nodes. For example, paths from PeriFW to a server or an endpoint can be identified. However, there are also physical paths that connect a server with an endpoint (via IDS and DMZRouter).

On the other hand, security policies such as access control policies, firewall policies, and network segmentation policies define the logical edges or paths that are subsets of the physical connections and paths. For example, an access policy that limits access to the file server only for endpoints with specific IDs enforces a logical path on the physical path from PeriFw to FileServer. Similarly, a firewall policy rule can limit remote access to FileServer only to authorised VPN traffic. Finally, the network segmentation policy can isolate the server pool from other network segments, allowing only specific internal services and users to access the servers within the pool. This segmentation helps prevent lateral movement in case of a security breach.

To enable automated CRI calculations, in our framework, the TTP attack flows are specified in JSON format as recommended by MITRE [25] .

*Figure 4. A segment of an example TTP attack flow in JSON.*

Policies are specified in eXtensible Access Control Markup Language (XACML) [27], which is a standard for specifying access control policies but is also suitable for firewall and network segmentation policies.

*Figure 5. An excerpt of an access control policy in XACML.*

Finally, a computer network/system G = ((V, I), E) is defined in GraphML[1], which is an XML-based format specifically designed for representing graphs. It is designed for graphs with built-in semantics for nodes, edges, and properties, and is suitable for complex graphs with many attributes and relationships.

*Figure 6. An example specification of a network graph in GraphML.*

## V. The CRI Calculation Procedure

Upon receiving the four inputs (AF, G, TI, Po) above for a set of attack flows, network graph, threat intelligence data, and set of security policies, the CRI calculation is completed as follows:

1. The nodes in an attack flow $af \in$ AF represent Tactics, Techniques, or Procedures (we call them TTP nodes). Some nodes can be the so-called STIX object that defines the supporting information such as involved asset, location, URL, etc.

2. For each attack flow in AF ( $af \in$ AF), for each TTP node $N$ in $af$, starting with the top node, we estimate how likely the network (G) with the security policies (Po) could be successfully attacked, given the threat intelligence data (PI). As a result, we will get a probability value, denoted by $P_N$.

3. If in $af \in AF$, any two nodes $N_1$ and $N_2$ are in a sequential or AND relation, then the cumulative probability of the network being successfully attacked is $P_{N_1} * P_{N_2}$.

4. If in $af \in AF$, two nodes $N_1$ and $N_2$ are in an OR relation, then the cumulative probability of being successfully attacked is $Max(P_{N_1}, P_{N_2})$.

5. The algorithm traverses the attack flow, and estimates the attack probability at each TTP node, and we denote the cumulative CRI for $af \in AF$ by CRI($af$).

6. The CRI value for the attack campaign is the maximum CRI or all attack flows in AF: $Max(CRI(af1), …, CRI(afn))$.

Further on point 2 above, at each TTP node of a MITRE attack flow, each technique (belongs to a tactic) can be modelled by a set of abstract attack trees, where each tree defines a high-level procedure. For example, within the Initial Access tactic (TA0001), the technique of DNS injection (T1659) can be modelled by the abstract attack tree in Figure 7. The tree is abstract because it does not specify the exact CVE vulnerabilities, but only the fact that it is a set of software vulnerabilities (CVE-RELATED). Besides it also contains the most typical attack steps of this technique.

---

[1] The GraphML File Format, http://graphml.graphdrawing.org/

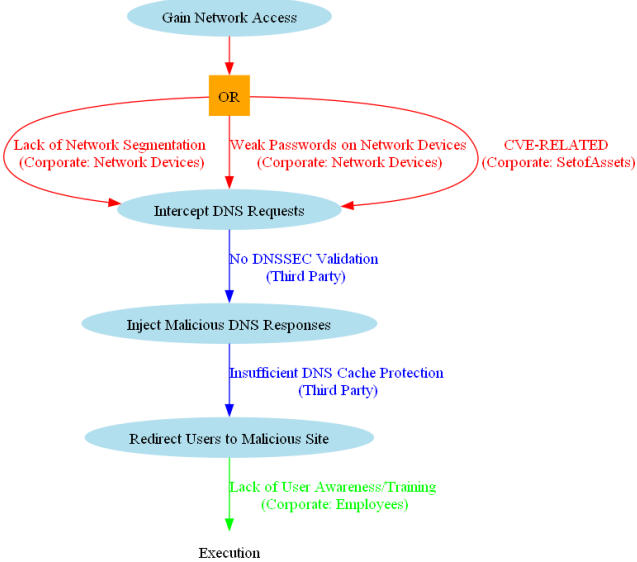

*Figure 7. An abstract attack tree for a type of DNS injection technique.*

For calculating the probability value of a successful attack ($P_N$), we rely on an approach that simulates and models the realistic decision-making behaviour of the attacker(s) against a fixed computer network/system, as follows:

1. We model the behaviour of the attacker using Partially Observable Markov Decision Process (POMDP). POMDP is a mathematical framework used to model decision-making in situations where the outcomes are partly random and partly under the control of a decision-maker (attacker), and where the full state of the system is not completely observable for the attacker. For example, after completing an attack step on the network, the attacker may observe the outcome as failed or successful. This makes it particularly suitable for modelling the behaviour of external attackers against a computer network. The behaviour of the attacker is guided by the TTPs known for an attack campaign, but the decision of the attacker to choose the next action in a TTP depends on the observed outcome of the previous action, the cost to carry out the next action, and the available historical success rate (if any).

2. We run an automated simulation of the attack attempts using the POMDP framework, given the input attack flow (*af*), the computer network (G), and security policies (Po), with the available threat intelligence data (TI).

3. For this, we define the states, actions, observations, and rewards based on the network and the input set of TTPs (capturing a cyber-attack campaign). Then, we use this POMDP model to simulate attack sequences and defensive responses, thereby enabling automatic attack simulations.

4. The probability value $P_N$ is calculated as the result of the state transitions in the Partially Observable Markov Decision Process.

In our POMDP model, the attacker(s) makes decisions to carry out the next steps of in the attack process based on the input TTP attack flow and the threat intelligence information (TI) such as the success rate in the past against a similar system, as well as the partial observation about the result of the previous attack step. The attacker can observe success or failure after an attack step, depending on the input computer network (G) and security policies in place (Po), and decide on the next step based on this observation and its threat intelligence data. A positive reward or negative reward can be defined for each step depending on the success or failure. For the implementation of our approach, we used Python and POMDPy [28] to manage the complexity of solving the POMDP model.

## VI. Mathematical Modelling

Following the high-level overview of the CRI calculation process provided in the previous section, in this section, we detail the mathematical modelling used in the process. The attacker(s) behaviour is modelled by Partially Observable Markov Decision Process, where we define the following elements:

- **States (S):** Different network states representing the security posture.
- **Actions (A):** Possible actions the attacker(s) took.
- **Observations (O):** Observations made by the attacker(s) after taking actions.
- **Transition Probabilities (T):** Probability of transitioning from one state to another given an action.
- **Observation Probabilities (O):** Probability of making a particular observation given a state and action.
- **Rewards (R):** Rewards associated with state-action pairs.
- **Initial Belief State (b0):** Initial probability distribution over the states.

In our case, states can be defined as related to the state of the computer system (e.g., nodes). Some example states related to the network in Figure 2 are as follows:

- s0: Initial state, no nodes compromised.
- s1: The attacker gets an initial foothold in the employee's remote endpoint.
- s2: Employee remote endpoint is compromised.
- s3: Email server is compromised.
- s4: The file server is compromised.
- s5: Privileges escalated on the user remote endpoint.
- s6: High-value target identified.
- s7: Moved laterally to the file server.
- s8: Data exfiltrated from the file server.

We also define the actions related to the MITRE ATT&CK Tactics, as follows:

- a1: Send phishing email.
- a2: Execute malicious attachment.
- a3: Install malware.

- a4: Establish persistence.
- a5: Escalate privileges.
- a6: Conduct internal reconnaissance.
- a7: Dump credentials.
- a8: Use obtained credentials.
- a9: Move laterally.
- a10: Identify high-value targets.
- a11: Exploit high-value target.
- a12: Exfiltrate data.

Besides the tactics related actions above, the actions are also defined for all the techniques and procedures. For example, the action "Inject Malicious DNS Response" in the abstract attack trees in Figure 7.

Since an external attacker in real life does not often see entirely the consequence of their action on the system, in this context, we define the following most typical high-level observations the attacker can make:

- o1: Success.
- o2: Failure.
- o3: Blocked.
- o4: Rejected.
- o5: Delayed response.
- o6: Access denied.
- o7: No response.
- o8: Error message.

***Transition Probabilities (T):*** In this context, the transition from a state $s$ to the next state $s'$ models the decision an attacker makes, and it is equal to the probability of selecting $s'$, given the current state $s$ and action $a$.

$$T(s, a, s') = P(s'|s, a)$$

For instance, $T(s0, a1, s1) = P(s1|s0, a1)$ is the probability of transitioning from the state "no nodes compromised" to the state "The attacker gets an initial foothold in the employee's remote endpoint" after sending the phishing.

***Observation Probabilities (O):*** Observation is an integral element of POMDP, in this context, it captures the probability of making an observation $o$ given that the system is in a state $s'$ and action $a$ was taken.

$$O(o|s', a) = P(o|s', a)$$

Considering the above transition, $O(o1|s1, a1) = P(o1|s1, a1)$ is the probability that the attacker makes an observation of success after sending the email. In our method, the probability of the observation made by the attacker(s) after each action is based on the input computer network (graph).

***Reward (R):*** Besides the observation made after each transition between states, the attacker is given a reward that can be negative if the step is unsuccessful and positive otherwise. The reward function is defined as $S \times A \rightarrow \mathbb{R}$, and denoted by R(s, a).

Rewards are crucial in determining the optimal actions within a POMDP framework as they guide the attacker's decision-making, where the attacker selects actions that maximize their expected reward.

***Policy (Strategy of the attackers):*** Over time, rewards help in developing a strategy that defines the best action to take in each belief state to achieve the highest cumulative reward.

***Initial Belief State (b0):*** The belief element is critical in the resilience index calculation within the POMDP framework because it represents the attacker's knowledge about the state of the network at any given time. In real-world scenarios, attackers often do not have complete knowledge about the network's state. They may only have indirect observations about the success or failure of their actions. The belief state provides a probabilistic representation of the network's state based on the attacker's observations and prior knowledge (including prior actions and observations). The initial belief state $b0(s)$ represents the initial probability distribution over states based on prior knowledge:

$$b0(s) = P(s)$$

For example: $b0(s0) = 1$, says that the initial belief that no nodes are compromised is 100%, while $b0(s0) = 0$ that initially the employee's remote endpoint is compromised is 0%.

***Belief Updates:*** The belief state is dynamically updated as the attacker takes actions and receives observations. This update reflects how the attacker's understanding of the network evolves over time, which is crucial for making decisions about subsequent actions under uncertainty. The belief state $b(s)$ is updated based on the action taken and the observation received:

$$b'(s') = wO(o|s', a)\sum T(s', s, a)b(s).$$

Based on this mathematical model, the pseudocode of the naive CRI calculation process is given as follows:

```
1. Input: Read Network Graph from GraphML File
2. Input: Read TTPs from JSON Files
3. Input: Security Policies from XCAML files
4. Input: Collect Historical Data (numerical values)

5. Generate the Set of States Based on the Network Graph

6. Estimate Transition and Observation Probabilities based
on threat intelligence data and historical statistics.

7. Estimate and Define the Rewards and Amounts.

8. Initialize Belief States.

9. Initialize the Cyber Resilience Index.

10. Explore All Possible Sequences of Actions
    a. For Each Possible Sequence of Actions in TTPs
        i. Initialize Current Belief State
        ii. Initialize Cumulative Reward for This Sequence
        iii. For Each Action in the Sequence
            - For Each Possible Observation
                - Update Belief State Using Transition and
Observation Probabilities
                - Update Cumulative Reward
```

```
            - Record the Action, Observation, Updated
    Belief State, and Cumulative Reward

    11. Calculate the Cyber Resilience Index (CRI)
        a. Normalize the Cumulative Reward
        b. Aggregate the Rewards to Form the Resilience Index

    12. Return CRI.
```

*Table 1. The naïve algorithm for calculating the CRI.*

## VII. Complexity

It is known that while POMDP provides a powerful framework for modelling complex decision-making problems under uncertainty, the challenges of the curse of dimensionality and state explosion are significant in the context of complex domains such as cybersecurity.

***Worst-case scenario****:* In the worst-case scenario, the complexity of a state transition with observation, when considering the sequential exploration of new states based on an action-state pair and then linking all possible observations to each new state, can be given by:

$$\text{Comp\_state\_obs} = |S| \times |A| \times |S| \times |O|$$

Where $|S|$ is the size of the state space (number of states), $|A|$ is the size of the action space, and $|O|$ is the size of the observation space. The state space depends on the number of nodes in the network graph (V), and the inventory set attached to the node (I). Since for each node, any combinations of the elements in the inventory set can be compromised, the state space can be calculated as:

$$|S| = (2^{|I|} - 1)\char`^|V|$$

where $|I|$ is the size of the largest inventory set in the network and $|V|$ is the number of nodes in the network.

The size of the action space can be $|A|$ is captured by the number of actions defined in the TTP Attack Flows. Namely, if we have the input attack campaign $AT = \{TTP1, ..., TTPn\}$, then $|A| = |TTP1| + \cdots + |TTPn|$, where $|TTP|$ is the size of the TTP attack flow.

Finally, the observation space is defined as the number of all possible observations in the given network. Some most typical observations are given in Section VI above.

Transition Matrix Complexity: The complexity of storing and computing the transition probabilities.

$$C_{statetrans} = |S| \times |A| \times |S|$$

For each action and each new state pair, $(a, s')$, as a result of the state transition above, we also store and assign all the possible observations, therefore ultimately, we get $|S| \times |A| \times |S| \times |O|$.

The complexity of calculating the optimal policy: In our POMDP model, the optimal policy calculation (if exists) is for determining the best actions to take by the attacker(s) in each state to maximize their expected reward over time. To determine the optimal policy, we rely on the so-called Bellman Update Equation, and the computation complexity will now be extended with the number of iterations until convergence.

Once we calculate the cumulative reward based on the policy of the attacker(s) against the given network based on the actions and decisions they make. The cyber resilience index can be calculated as:

$$\text{CRI} = \frac{\varSigma Ri}{Number\ of\ Simulations}$$

where $\varSigma Ri$ is the cumulative rewards following the optimal policy over several simulation rounds. To increase the accuracy of the CRI value, we generate a representative set of simulation scenarios. During each simulation (scenario), the attacker follows the optimal policy. Each simulation may end up with a different value of cumulative rewards.

***Reducing the complexity****:* One can see that in the worst-case scenario, the complexity is exponential, and therefore it takes a long time to calculate CRI (in general).

However, we managed to reduce greatly the complexity using the following approaches:

1.  To reduce the state space, we complete a pre-processing of the input set of attack flows, and for each TTP node in an attack flow, we select only the subset of nodes in the network that are likely to be the target based on the related tactic, technique, procedure. With the same approach, we can reduce the size of the inventory set by focusing only on the ones that are likely to be targeted based on the related TTP and threat intelligence.

2.  Because in our case, the decisions of the attacker(s) are guided by the TTP nodes in the attack flow, the state transition is not arbitrary as after a given state only a limited set of actions will be selected, and it will only transit to a limited set of new states. Therefore, $C_{statetrans}$ can be greatly reduced.

3.  The number of observations applicable to each pair of actions and state can also be reduced greatly compared to the worst-case scenario. Instead of considering all observations in the observation space, we can narrow it down only to a subset of observations that are the typical known outcomes of an action specified in a given TTP against the target node. For this, given a target node in the network topology, before solving the POMDP, we generate the corresponding possible observations as a result of the action on the nodes along the physical and logical paths from the network perimeter (e.g., public-facing nodes) to the target node.

To summarize, the updated algorithm is given as follows:

```
1. Input: Read Network Graph from GraphML File
2. Input: Read TTPs from JSON Files
3. Input: Security Policies from XCAML files
4. Input: Collect Historical Data (numerical values)

5. Pre-process Attack Flows
     a. For each TTP Node in an Attack Flow
          i. Identify Likely Target Nodes in the Network
               - Use TTP-related tactics, techniques, and
procedures
               - Select a subset of nodes based on their
likelihood of being targeted
          ii. Reduce Inventory Set
               - Focus only on items likely to be targeted
based on TTP and threat intelligence

6. Generate the Set of States Based on the Reduced
Network Graph
     a. Consider only nodes identified in pre-processing
     b. Use reduced inventory set

7. Estimate Transition and Observation Probabilities
     a. Use historical statistics and threat intelligence
     b. Limit transitions to relevant TTP-guided actions
          - Transition is not arbitrary; only limited sets
of actions are considered
          - Only transitions to a limited set of new
states are allowed

8. Estimate and Define the Rewards and Amounts

9. Initialize Belief States

10. Initialize the Cyber Resilience Index

11. Explore Feasible Sequences of Actions
     a. For Each Possible Sequence of Actions in TTPs
          i. Initialize Current Belief State
          ii. Initialize Cumulative Reward for This
Sequence
          iii. For Each Action in the Sequence
               - Identify the Set of Applicable
Observations
               - Only typical known outcomes of the
action are considered
               - Generate observations based on the
physical and logical paths to the target node
               - For Each Possible Observation
                    - Update Belief State Using Transition
and Observation Probabilities
                    - Update Cumulative Reward
                    - Record the Action, Observation,
Updated Belief State, and Cumulative Reward

12. Calculate the Cyber Resilience Index (CRI)
     a. Normalize the Cumulative Rewards
     b. Aggregate the Rewards to Form the Resilience
Index

13. Return CRI.
```

*Table 2. The proposed improved algorithm for calculating the CRI.*

## VIII. CRI Outputs & Discussion

The Cyber Resilience Index (CRI) needs to be well understood by decision-makers, thereby it serves as an aggregated metric that provides an overview of how well an organisation's security posture stands against threats and their corresponding threat actors. This metric is continuously reassessed and reevaluated. Organisations should understand their applicable threat landscape and relevant threat actors. An example of the CRI for ransomware threats is shown in Figure 7.

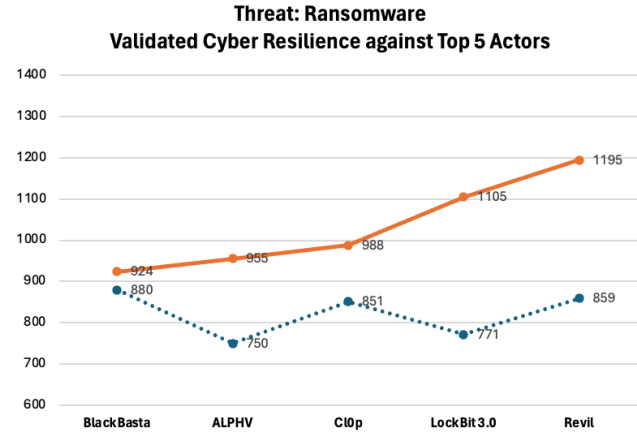

*Figure 8. Validated (orange) versus (assumed) cyber resilience against top 5 ransomware threat actors.*

The blue line indicates the assumed cyber resilience before the simulation or validation of the security controls in place. The orange line represents the validated cyber resilience against the top 5 ransomware-related threat actors after emulating their modus operandi. To account for uncertainty, the most probable attack path deviating from the CTI-given modus operandi is included in the scope of emulation and our calculations. This is a key differentiator compared to static threat-informed defences evaluated purely against explicit CTI-provided inputs.

The cyber resilience index follows the principles of a stock market index. Specifically, when a new threat actor, such as BlackBasta, is identified as applicable to an organisation, a dip in the assumed defence against its modus operandi can be observed. However, after careful evaluation and emulation, the overall cyber resilience should increase, unless risks are accepted or treated in a way that justifies the index's reading.

A supplementary metric used to assist decision-makers and teams is the confidence graph in the effectiveness of security controls and countermeasures versus their capital, operational, and maintenance costs. We use MITRE D3fend [29] to match the TTPs of MITRE ATT&CK, adopting a security-driven approach rather than a compliance-driven one. However, organisations may choose to base their cost-benefit analysis of security controls on higher-level or compliance-driven frameworks such as NIST [30], the ISO 27000 series [31], or CIS [32]. An example is shown in Figure 8, using the countermeasure groups of D3fend: harden, detect, isolate, deceive, evict, and restore. As a result, decision-makers can make informed decisions on investing in and implementing new controls or countermeasures based on the added value to their overall CRI, which is grounded in risk tolerance and risk appetite.

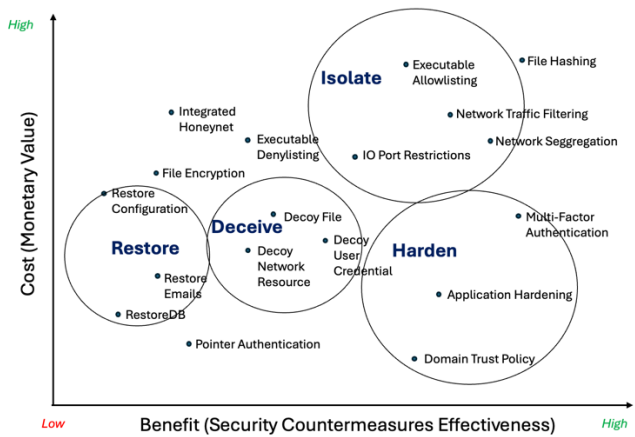

*Figure 9. Confidence graph in security control & countermeasure effectiveness based on cost vs benefit.*

## IX. Conclusions & Future Research

In the evolving landscape of cybersecurity, traditional risk assessment methodologies often fall short in addressing the dynamic nature of threats. This paper introduces the Cyber Resilience Index, an innovative, probabilistic approach to quantifying an organisation's security posture. By integrating the Threat-Intelligence Based Security Assessment (TIBSA) methodology with CRI, we provide a unified, actionable metric akin to a stock market index. This enables decision-makers to make informed, data-driven decisions about their cybersecurity investments and strategies, enhancing their ability to respond to emerging threats effectively. The CRI is a practical tool for continuous improvement, bridging the gap between qualitative and quantitative assessments, allowing for a more resilient cyber environment backed by factual accuracy rather than assumed protective measures or theoretical compliance-driven checklists.

With the introduction of CRI, several future research directions are opened. For instance, how to further refine the CRI framework with the use of advanced predictive analytics. Full automation of CRI calculation is another very promising research direction. Developing an automated tool to streamline the CRI calculation process, making it more accessible and user-friendly for organisations of all sizes. Continuously updating and expanding the libraries of attack flows and TTPs to reflect the latest threat landscapes, while being able to predict the adversary's next moves based on the historical datasets, is another potential research direction. This would reduce inherent uncertainty while adversaries are slightly deviating from their observed modus operandi subject to the uniqueness of each network.

Establishing benchmarks across different industries to provide comparative insights and best practices for enhancing cyber resilience, and therefore collectively raising the cyber resilience bar as well as the know-how amongst sectors would be a great future research direction. Lastly, incorporating feedback from end-users to refine the CRI framework, ensuring it meets the practical needs of security teams and decision-makers, would help build a common language for measuring effectiveness and efficiency on top of the well-established threat-informed defence concepts.